\documentclass[a4paper]{jpconf} 
\usepackage{amsmath}
\usepackage{amsthm} 
\usepackage{graphicx}
\usepackage{dcolumn}
\usepackage{bm}
\usepackage{amsfonts}

\usepackage{setspace}
\usepackage{braket}
\usepackage{graphics}

\begin{document}
\title{Breakdown of the Meissner effect at the zero exceptional point in non-Hermitian two-band BCS model}

\author{Takano Taira}

\address{Research Fellow of Japan Society for the Promotion of Science, \\Institute of Industrial Science, The University of Tokyo
\\5-1-5 Kashiwanoha, Kashiwa 277-8574, Japan}

\ead{taira904@iis.u-tokyo.ac.jp}

\begin{abstract}
    The spontaneous symmetry breaking of a continuous symmetry in complex field theory at the exceptional point of the parameter space is known to exhibit interesting phenomena, such as the breakdown of a Higgs mechanism. In this work, we derive the complex Ginzburg-Landau model from a non-Hermitian two-band BCS model via path integral and investigate its spontaneous symmetry breaking. We find that analog to the Higgs mechanism, the Meissner effect of the complex Ginzburg-Landau model also breaks down at the exceptional point while the gap parameters stay finite. 
\end{abstract}

\section{Introduction}
Non-Hermitian physics has been a subject of intense theoretical and experimental research over the past few decades, with one of its key features being the eigenvalue parameter point known as the exceptional point (EP) \cite{kato2013perturbation}. At this point, the eigenvalues and eigenvectors of the Hamiltonian coalesce, reducing the rank of the Hamiltonian. The number of coalescing eigenvectors determines the order of the EP, with the most common type of the exceptional point being the order 2 EP (EP2), where two eigensystems coalesce, and eigenvalues become complex conjugate pairs beyond the EP2. Additionally, the branch-cut structure of the EP causes eigenvector switching when parameters of the Hamiltonian are varied around the EP.  In our prior investigations into the non-Hermitian Higgs mechanism \cite{fring2020goldstone,fring2022massive}, we have found another kind of EP2. In this variant, one eigenvalue becomes zero, while another eigenvalue remains real even after crossing the EP, deviating from the typical branch-cut structure associated with EPs. We have termed this point the zero exceptional point (0EP), distinguishing it from the conventional EP2.

The EP of the non-Hermitian system has important implications in systems where the Hamiltonian and its eigenvectors exhibit anti-linear symmetry, such as parity time-reversal ($\mathcal{PT}$) symmetry. In these scenarios, the anti-linear symmetry of the eigenvector breaks down at the EP, leading to the emergence of complex conjugate pairs of eigenvalues \cite{bender1998real}.  The EPs have been observed in a variety of physical phenomena, such as the stopping of light in classical optics \cite{goldzak2018light}, quantum phase transitions \cite{heiss2005large,yamamoto2019theory,hanai2019non}, and more examples to be found in the review paper \cite{heiss2012physics}. Despite significant research on EPs in non-Hermitian one-body systems, their analysis in non-Hermitian many-body systems is still in its early stages both theoretically and experimentally \cite{durr2009lieb,ashida2016quantum,nakagawa2018non,lourencco2018kondo,daley2009atomic,begun2021phase,kosov2011lindblad,ghatak2018theory,yamamoto2019theory,kantian2009atomic,garcia2009dissipation,kozii2017non,hanai2020critical,hanai2019non}. 

\begin{figure}[h]
            \centering
            \begin{minipage}[b]{0.4\textwidth}           \includegraphics[width=7cm]{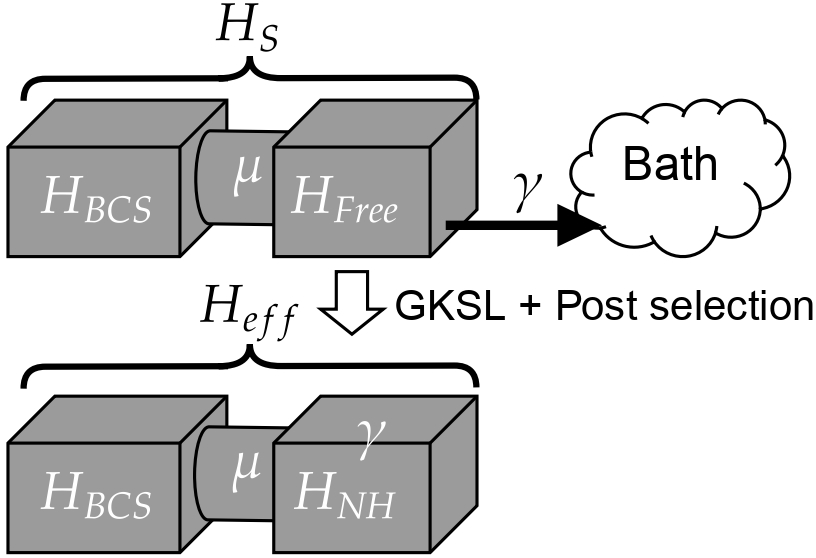}
            \end{minipage}
            \caption{Schemetic of the system Hamiltonian $H_S$, coupled to the external bath. The large vertical arrow indicates that we have made an appropriate approximation to reduce the system to a non-Hermitian effective Hamiltonian. The bath coupling parameter $\gamma$ becomes one of the parameters in the effective Hamiltonian $H_\text{eff}$ after GKSL and post-selection.}
            \label{figure : schematic}
        \end{figure} 

One of the most notable achievements of the many-body system is its success in explaining the complex phenomena of superconductivity. A superconductor is characterized by two key macroscopic phenomena: zero electrical resistivity and diamagnetism (Meissner effect). These phenomena are described by the first and second London equations, respectively \cite{london1935electromagnetic}. Specifically, the equations are $\nabla^2 \vec{B} \propto n \vec{B}$ and $\partial_t \vec{j} \propto n \vec{E}$, where $\vec{B}$, $\vec{j}$, and $\vec{E}$ are the magnetic field, current density, and electric field, respectively. The constant $n$ is proportional to the absolute value of the gap parameter from the Bardeen–Cooper–Schrieffer (BCS) theory \cite{bardeen1957theory}. Therefore, if the gap parameter is zero, the Meissner effect and zero electrical resistivity phenomena are disrupted, resulting in a breakdown of superconductivity. 

In this paper, we build upon our earlier findings on the breakdown of the Higgs mechanism at the 0EP in a non-Hermitian scalar field theory \cite{fring2020goldstone,fring2022massive}. Our prior analysis revealed that the mechanism behind this breakdown was the spontaneous symmetry breaking of both the anti-linear ($\mathcal{PT}$) symmetry and the continuous $SU(2)$ symmetry. Extending this framework, we now focus on a non-Hermitian many-body model, demonstrating its correspondence with the non-Hermitian scalar field theory via path-integral formalism. The analog of the Higgs mechanism in this model would be the Meissner effect. The resulting non-Hermitian scalar field theory has continuous $U(1)$ symmetry with 0EP in the parameter space. However, our model differs from our prior work \cite{mannheim2019goldstone,fring2022massive} where no obvious anti-linear symmetry is associated with the 0EP. Nonetheless, we find the Meissner effect breaks down at the 0EP.

This paper is organized as follows. In section 2, we begin with the Hermitian many-body model in the context of the open quantum system. We assume that our model can be approximated by the Gorini-Kossakowski-Sudarshan-Lindblad (GKSL) master equation \cite{gorini1976completely,lindblad1976generators} to obtain the effective non-Hermitian many-body Hamiltonian. We arrive at the non-Hermitian scalar field theory using the path-integral formalism and the non-Hermitian mean field theory \cite{ashida2016quantum}. Section 3 calculates the non-Hermitian London equation from the non-Hermitian scalar field theory obtained in section 2 and investigates the parameter point where the Meissner effect breaks down. Finally, we show that such a parameter point corresponds to the 0EP of the non-Hermitian scalar field theory, arriving at the main message of this paper: \textit{The Meissner effect breaks down at the 0EP of the non-Hermitian scalar field theory.}

\section{The two-component complex Ginzburg-Landau model}\label{Section: The non-Hermitian two-band BCS model}
\subsection{Many-body Hamiltonian}
We aim to find a many-body Hamiltonian that admits the non-Hermitian scalar field theory with continuous symmetry, similar to the models previously investigated in the context of Higgs mechanism \cite{alexandre2017symmetries,mannheim2019goldstone,fring2022massive}. We begin with the BCS and Free Hamiltonians of different bands, coupled via two-band superconductor-type coupling \cite{suhl1959bardeen}, which is schematically depicted in Fig. \ref{figure : schematic}. The system Hamiltonian is $H_{\text{S}} = H_{\text{BCS}}+H_{\text{Free}}+H_{\text{Int}}$ where individual Hamiltonians are 
\begin{eqnarray}
        H_{\text{BCS}} &=&\int d^3 r \sum_{\sigma=\uparrow , \downarrow}~ c^\dagger_{1\sigma} (\vec{r})\left(-\frac{\nabla^2_r}{2m_1 }-\mu_1\right)c_{1\sigma} (\vec{r})- g c^\dagger_{1\uparrow} (\vec{r})c^\dagger_{1\downarrow} (\vec{r})c_{1\downarrow} (\vec{r})c_{1\uparrow} (\vec{r}),\label{BCS Hamiltonian}\\
        H_{\text{Free}}&=&\int d^3 r \sum_{\sigma=\uparrow , \downarrow}~ c^\dagger_{2\sigma} (\vec{r}) \left(-\frac{\nabla^2_r}{2m_2 }-\mu_2\right) c_{2\sigma} (\vec{r}),\label{Free Hamiltonian}\\
        H_{\text{int}}&=&- \mu \int d^3 r ~ c_{1\uparrow}^\dagger (\vec{r})c_{1\downarrow}^\dagger (\vec{r})c_{2\uparrow}(\vec{r})c_{2\downarrow}(\vec{r})+c_{2\uparrow}^\dagger (\vec{r})c_{2\downarrow}^\dagger(\vec{r}) c_{1\uparrow}(\vec{r})c_{1\downarrow}(\vec{r})(\vec{r}).\label{Interaction Hamiltonian}
\end{eqnarray}
The creation and the annihilation operators of the two bands are given by $\{c_{1\sigma}^\dagger ,c_{1\sigma}\}$ and $\{c_{2\sigma}^\dagger ,c_{2\sigma}\}$, where $\sigma=\uparrow, \downarrow$. The three coupling constants $g$, $\gamma$, and $\mu$ correspond to the internal coupling of BCS Hamiltonian, the coupling between the free Hamiltonian and the bath Hamiltonian, and the coupling between BCS Hamiltonian and free Hamiltonian, respectively, all of which are assumed to be positive. The constants $e $, $\{m_1,m_2\}$ and $\{\mu_1,\mu_2\}$  are the gauge charge, the unnormalized masses of the particle for each band, and the chemical potential for each band, respectively. {We note that a similar model was considered in Ref. \cite{yamamoto2021collective} from the perspective of the fermionic superfluid.}

The time evolution of the system's density matrix is described by the Gorini-Kossakowski-Sudarshan-Lindblad (GKSL) master equation \cite{gorini1976completely,lindblad1976generators}, given by 
\begin{eqnarray}
    \frac{d \rho}{dt} =  -i (H_{\text{NH}}\rho - \rho H_{\text{NH}}^\dagger)- \frac{i}{\hbar} \left[ H_{\text{Lamb}},\rho\right]+ i\gamma \int d^3 r  L_2 (\vec{r}) \rho L_2^\dagger (\vec{r}),\nonumber
\end{eqnarray}
where $H_{\text{NH}}$ is the effective non-Hermitian Hamiltonian defined as $H_{\text{NH}} = H_{\text{free}} - i\gamma/2 \int d^3r L_2^\dagger L_2$. The Lindblad operators $L_2(\vec{r}) = c_{2\downarrow} (\vec{r})c_{2\uparrow}(\vec{r})$ and $L^\dagger_2 (\vec{r})  =  c_{2\uparrow}^\dagger (\vec{r}) c_{2\downarrow}^\dagger (\vec{r})$ create and destroy the two-particle state in the system at a rate of $1/\gamma$. {Such two-body loss in BCS superconductor was previously considered in Ref. \cite{kosov2011lindblad} and recent works have discussed the possibility of two-body losses in the solid-state setup \cite{visuri2023dc,huang2023modeling}.} The Lamb-shift term $H_{\text{Lamb}}$ introduces a real shift in the energy level of the free Hamiltonian $H_{\text{free}}$. 

{This GKSL master equation approximately captures the dynamics of the system when the coupling between the system and the bath is small, and the bath is large enough to maintain the thermal equilibrium (Born-Markov approximation)}\cite{breuer2002theory}. Furthermore, {the dynamics of the system in a short time scale is approximately dictated by the non-Hermitian Hamiltonian } $d \rho/d t =  -i (H_{\text{NH}}\rho - \rho H_{\text{NH}}^\dagger)$, omitting the last two terms of the above dynamical equation. This is called a post-selection, often used in open quantum systems to obtain the non-Hermitian Hamiltonian  \cite{naghiloo2019quantum}. {Put simply, our model essentially functions as a standard two-band BCS model with a perturbative effect due to the bath. A simpler setup was previously considered in Ref. \cite{kosov2011lindblad} but did not consider the Meissner effect.} The effective non-Hermitian Hamiltonian, which is depicted schematically in Fig. \ref{figure : schematic}, is given by $H_{\text{eff}}=H_{\text{BCS}}+H_{\text{Int}}+H_{\text{NH}}$. 

As a supplementary observation, we mention that the non-Hermitian effective Hamiltonian $H_{\text{NH}} [a^\dagger_2,a_2]$ could also be manifested as an effective theory of an ultra-cold atomic gas \cite{yamamoto2019theory,durr2009lieb,liu2022weakly,yamamoto2021collective} with recent experimental progress \cite{takasu2020pt}. {Furthermore, a small body of work has investigated the Meissner effect in ultra-cold atoms \cite{atala2014observation,cano2008meissner}.} This correspondence provides a promising avenue for exploring non-Hermitian physics in a controllable laboratory setting.

\subsection{non-Hermitian scalar field theory}
    To find the non-Hermitian scalar field theory from the many-body Hamiltonian form previous section, we begin our analysis from the partition function $Z =\int \mathcal{D} \psi_1  \mathcal{D} \psi_1^\dagger  \mathcal{D} \psi_2\mathcal{D} \psi_2^\dagger \exp(- S)$ where the complex action $S$ is given by rewriting the effective Hamiltonian $H_{\text{eff}}$ in terms of coherent-state path integral. The two sets of Grassmannian fields $\{\psi_{1\sigma} ,\psi_{1\sigma}^\dagger\}$ and $\{\psi_{2\sigma} ,\psi_{2\sigma}^\dagger\}$, $\sigma = \uparrow, \downarrow$, correspond to the coherent states of the creation and the annihilation operators $\{c_{1\sigma} , c_{1\sigma}^\dagger\}$ and  $\{c_{2\sigma} , c_{2\sigma}^\dagger\}$, respectively. The explicit form of the action can be found by following the standard procedure of condensed matter theory \cite{breuer2002theory}, and it is given as
    \begin{eqnarray}
        S&=& S_{1}[\psi_1]+S_{2}[\psi_2] + S_{\text{Int}}[\psi_1,\psi_2],\label{Total action}\\
        S_{1} &=& \sum_{\sigma=\uparrow \downarrow}\int_0^{1/T} d \tau\int d^3 r  \psi^\dagger_{1\sigma} (\vec{r})  \left(\partial_\tau-\frac{\nabla_r^2}{2m_1} - \mu_1 \right) \psi_{1\sigma} (\vec{r})-g \psi^\dagger_{1\uparrow }(\vec{r}) \psi^\dagger_{1\downarrow }(\vec{r}) \psi_{1\downarrow }(\vec{r}) \psi_{1\uparrow }(\vec{r}),\label{BCS action}\\
        S_{2} &=& \sum_{\sigma=\uparrow \downarrow}\int_0^{1/T} d \tau\int d^3 r \psi^\dagger_{2\sigma} (\vec{r}) \left(\partial_\tau-\frac{\nabla_r^2}{2m_2} - \mu_2\right) \psi_{2\sigma} (\vec{r})- i\gamma  \psi^\dagger_{2\uparrow }(\vec{r}) \psi^\dagger_{2\downarrow }(\vec{r}) \psi_{2\downarrow }(\vec{r}) \psi_{2\uparrow }(\vec{r}),~~~~\\
        S_{\text{Int}}&=&-\mu \int_0^{1/T} d \tau\int d^3 r~\psi^\dagger_{1\uparrow}(\vec{r})\psi^\dagger_{1\downarrow}(\vec{r})\psi_{2\uparrow}(\vec{r})\psi_{2\downarrow}(\vec{r})+\psi^\dagger_{2\uparrow}(\vec{r})\psi^\dagger_{2\downarrow}(\vec{r})\psi_{1\uparrow}(\vec{r})\psi_{\downarrow}(\vec{r}),\label{Coupling Hamiltonian}
        \end{eqnarray}
        where $\{\tau,\vec{r}\}$ are Wick roateted imaginary time $-i\tau = t$ and 3-dimensional space. The imaginary time is integrated between $0$ and the inverse temperature $1/T$, where $T$ is the temperature of the whole system. 
            
    Following the non-Hermitian mean field theory of Ref. \cite{yamamoto2019theory}, we consider the partition function of the auxiliary fields $\{\Delta_1 , \overline{\Delta}_1\}$ and $\{\Delta_2 , \overline{\Delta}_2\}$: $Z_\Delta =\prod_{i=1,2} \int D\overline{\Delta}_i  D\Delta_i \exp\big[- \frac{1}{g}\overline{\Delta}_1 \Delta_1 - \frac{1}{i \gamma} \overline{\Delta}_2 \Delta_2  \frac{  \mu}{i \gamma g }(\overline{\Delta}_1 \Delta_2 + \overline{\Delta}_2 \Delta_1)$, where $\overline{\Delta}_i$ is not necessarily a complex conjugate of $\Delta_i $ for $i=1,2$. Multiplying the partition function $Z$ by $Z_\Delta / Z_\Delta$ we find $Z \propto \prod_{i=1,2} \int D \psi_i^\dagger D\psi_i D \overline{\Delta}_i D \Delta_i \exp(-\Tilde{S})$, where $\Tilde{S} =S + \frac{1}{g}\overline{\Delta}_1 \Delta_1 + \frac{1}{i \gamma} \overline{\Delta}_2 \Delta_2 + \frac{\mu}{i \gamma g }(\overline{\Delta}_1 \Delta_2 + \overline{\Delta}_2 \Delta_1)$. Performing the Hubbard Stranovovich transformation
    \begin{eqnarray}
        \Delta_1 \rightarrow \Delta_1 - g \psi_{1\downarrow}\psi_{1\uparrow}, & \overline{\Delta}_1 \rightarrow \overline{\Delta}_1 - g \psi^\dagger_{1\uparrow}\psi_{1\downarrow}^\dagger ,~~~~\\
        \Delta_2 \rightarrow \Delta_2 - i \gamma \psi_{2\downarrow}\psi_{2\uparrow}, & \overline{\Delta}_2 \rightarrow \overline{\Delta}_2 - i\gamma \psi^\dagger_{2\uparrow}\psi_{2\downarrow}^\dagger,~~~~
    \end{eqnarray}
    and integrating out the Grassmannian fields, we find the effective action only in terms of the auxiliary bosonic fields:
    \begin{eqnarray}\label{effective action}
    	 S_{\text{eff}}&=&  -\text{Tr}\log \left(\begin{array}{cc}
				i\omega_n + \epsilon_{\vec{k}}^{(1)}&\Delta_1 + \frac{\mu}{i\gamma}\Delta_2 \\
				\overline{\Delta}_1 + \frac{\mu}{i\gamma}\overline{\Delta}_2&i\omega_n - \epsilon_{\vec{k}}^{(1)}
    	\end{array}\right)- \text{Tr}\log \left(\begin{array}{cc}
    	i\omega_n + \epsilon_{\vec{k}}^{(2)}&\Delta_2 + \frac{\mu}{g_1}\Delta_1 \\
    	\overline{\Delta}_2 + \frac{\mu}{g_1}\overline{\Delta}_1&i\omega_n - \epsilon_{\vec{k}}^{(2)}
    \end{array}\right)\nonumber\\
	&& + \int_0^{1/T} d \tau \int d^3 r\quad\frac{\mu}{i \gamma g}(\overline{\Delta}_1 \Delta_2 + \overline{\Delta}_2 \Delta_1)+ \frac{1}{g}\overline{\Delta}_1 \Delta_1 + \frac{1}{i\gamma}\overline{\Delta}_2 \Delta_2 , 
    \end{eqnarray} 
    where $\epsilon_r^{(i)} \equiv \nabla_r^2 / 2m_i + \mu_i$ for $i=1,2$. Solving the equations of motion $\delta S_{\text{eff}} / \delta \Delta_i =0 $, and $\delta S_{\text{eff}} / \delta \overline{\Delta}_i =0 $ for $i=1,2$ and defining the expectation value by the path-integral $\braket{\dots }\equiv \int \mathcal{D}\psi^\dagger \mathcal{D}\psi \dots e^{-S}$, one finds the gap equations 
    \begin{equation}
        \begin{array}{cc}
            \frac{1}{g} \left(\Delta_1 + \frac{\mu}{i\gamma}\Delta_2\right) =\braket{\psi_{1\downarrow}\psi_{1\uparrow}} + \frac{\mu}{g} \braket{\psi_{2\downarrow}\psi_{2\uparrow}}, & \frac{1}{g}\left(\overline{\Delta}_1 + \frac{\mu}{i\gamma}\overline{\Delta}_2\right) =\braket{\psi_{1\uparrow}^\dagger \psi_{1\downarrow}^\dagger} + \frac{\mu}{g} \braket{\psi_{2\uparrow}^\dagger \psi_{2\downarrow}^\dagger} \\
             \frac{1}{i\gamma} \left(\Delta_2 + \frac{\mu}{g}\Delta_1\right) 
            = \braket{\psi_{2\downarrow}\psi_{2\uparrow}} + \frac{\mu}{i\gamma} \braket{\psi_{1\downarrow}\psi_{1\uparrow}},& \frac{1}{i\gamma}\left(\overline{\Delta}_2 + \frac{\mu}{g}\overline{\Delta}_1\right) =\braket{\psi_{2\uparrow}^\dagger \psi_{2\downarrow}^\dagger } + \frac{\mu}{i\gamma} \braket{\psi_{1\uparrow}^\dagger\psi_{1\downarrow}^\dagger} \label{Eq. Gap equation}
        \end{array}
    \end{equation}
    Notice that the two-point functions such as $\braket{\psi_{1\downarrow}\psi_{1\uparrow}}$, depending on the temperature of the whole system $T$ and the parameters of the model $e, m_i, \mu_i, g ,\gamma ,\mu$ where $i=1,2$. Therefore, solving the above gap equations leads to the solution to the auxiliary fields $\{\Delta_i (T), \overline{\Delta}_i (T)\}$, $i=1,2$ where they also depend on the temperature $T$ and the model parameters. 
    
    Let us comment on the auxiliary bosonic fields. If we assume $\overline{\Delta}_i$ to be the complex conjugate of $\Delta_i$, then the gap equations (\ref{Eq. Gap equation}) can not be solved for $\Delta_i$. Such incompatibility of equations is a common problem in non-Hermitian systems \cite{mannheim2019goldstone,fring2020goldstone,fring2022massive}, where there are several ways to resolve this issue. Here, we will adopt the approach outlined in Ref. \cite{yamamoto2019theory} and leave $\overline{\Delta}_i$ independent of $\Delta_i$. 

    To investigate the Meissner effect of our model, we Taylor expand the trace-log term of the effective action (\ref{effective action}) to obtain the non-Hermitian scalar field theory in terms of the auxiliary fields. The expansion is most conveniently performed with respect to the combined fields $\Delta_\epsilon \equiv \Delta_1 + i \epsilon \Delta_2$, $\Delta_\delta \equiv \Delta_2 + \delta \Delta_1$, $\overline{\Delta}_\epsilon \equiv \overline{\Delta}_1 + i \epsilon \overline{\Delta}_2$ and $\overline{\Delta}_\delta \equiv \overline{\Delta}_2 + \delta \overline{\Delta}_1$, where $\epsilon = \mu / \gamma $ and $\delta = \mu / g$ are dimensionless parameters. However, solving the gap equations (\ref{Eq. Gap equation}) is necessary to justify the expansion of the effective action around the critical temperature where the combined fields vanish.
    \subsection{The gap equation}\label{Section: The gap equation}
        To obtain numerical solutions for the gap equations (\ref{Eq. Gap equation}), we employ the approximations $\epsilon \ll \delta \ll 1$ and truncate the equations up to, but not including, terms of order $\delta \epsilon$. {This hierarchy corresponds to $\gamma \gg g \gg \mu$.} It is worth noting that this approximation scheme has been chosen for illustrative purposes, and other alternatives remain for future investigation.

\begin{figure}
            \vspace{-10pt}
            \centering
            \begin{minipage}[b]{0.5\textwidth}           \includegraphics[width=\textwidth]{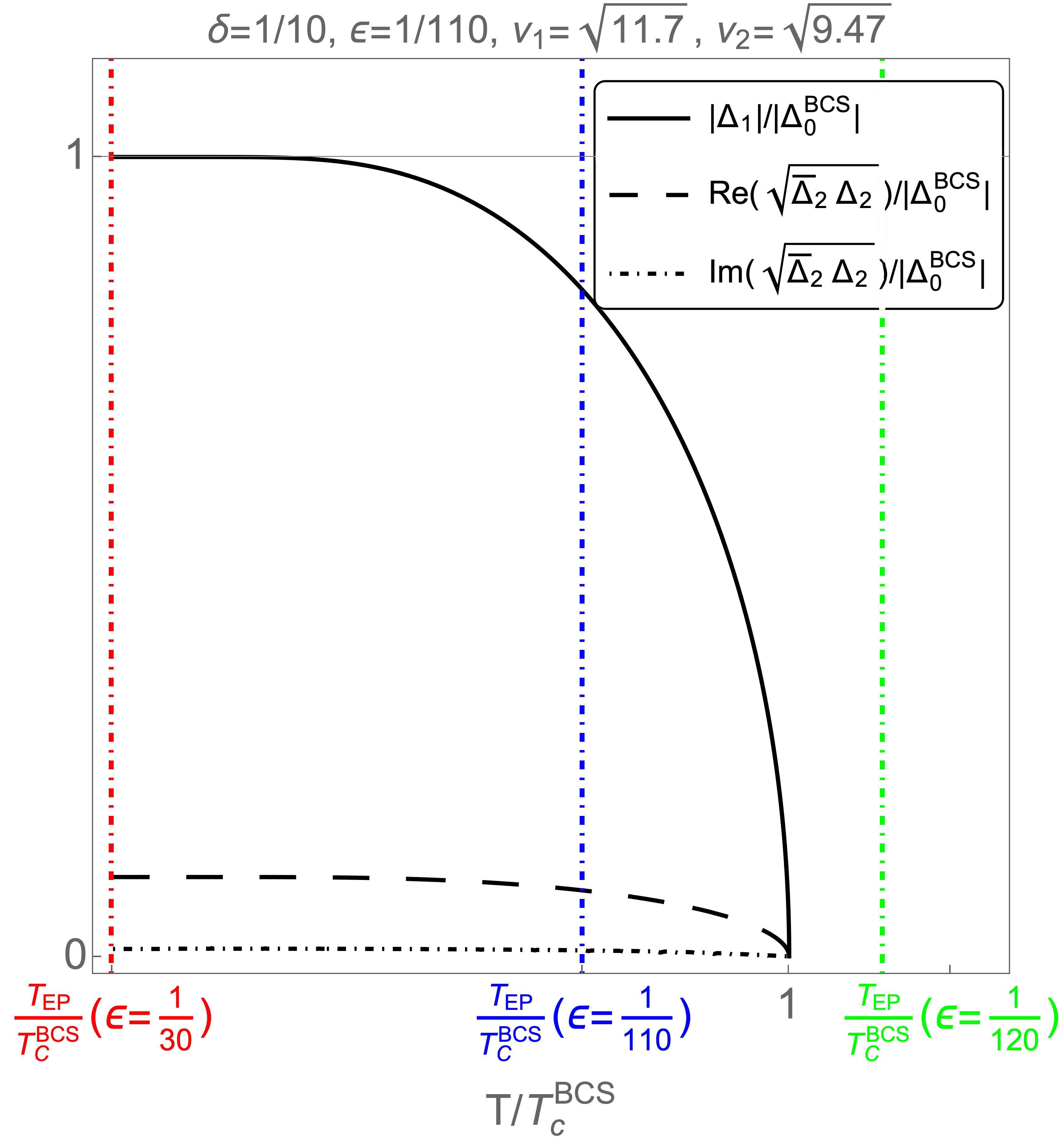}
            \end{minipage}
            \vspace{-15pt}
            \caption{The figure displays numerical solutions for the gap parameters as a function of temperature, obtained from the gap equations $\delta S_{\text{eff}}/ \delta \Delta_1 =0 $ and $\delta S_{\text{eff}}/ \delta \Delta_2 =0 $ with the approximation $\epsilon \ll \delta \ll 1$, neglecting terms of order $\delta \epsilon$. The parameters used are $1/g \nu_1 = 1/0.39$, $\nu_1 = \sqrt{11.7}$, $\nu_2 = \sqrt{9.47}$, and $\delta = 1/10$ with $\epsilon = 1/ 110$. The solid and dotted lines represent the real and imaginary parts of the solutions, respectively. The figure also includes three vertical lines corresponding to the exceptional temperatures for $\epsilon=1/30$, $\epsilon = 1/110$, and $\epsilon= 1/120$, respectively, where the Meissner effect breaks down.}
            \label{figure : gap solution}
            \vspace{-20pt}
        \end{figure}

        We numerically solve the simplified gap equation obtained using the above approximation. Details of the calculation are provided in the \ref{Appendix: HS transformation}, as this is not the main focus of this paper. The solutions are presented in Fig. \ref{figure : gap solution}, where we have taken Al (Aluminium) as a BCS superconductor with a density of states at the Fermi level of $\nu_1 = \sqrt{11.7}, \text{eV}$ and $1/g \nu_1 = 1/0.39$ as reported in Ref. \cite{de2018superconductivity}. For the free theory $H_0 [c_2 , c_2^\dagger]$, we use the density of states $\nu_2 = \sqrt{9.47}, \text{eV}$ of Zn (Zinc). We emphasize that these experimental values are chosen for illustrative purposes and that the experimental realization of our model is left for future research. Choosing different experimental values also produces a plot similar to Fig. \ref{figure : gap solution} with different critical temperatures, allowing a Taylor expansion of the effective action near the critical temperature. {We further note that the rest of the derivation of this manuscript does not depend on the experimental values chosen above, and our main result discussed in section 3 is independent of the experimental values chosen above.} 

        In Fig. \ref{figure : gap solution}, we plot the gap parameters $|\Delta_1(T) |/ \Delta^{\text{BCS}}_0$ and $\sqrt{\overline{\Delta}_2\Delta_2(T)}/ \Delta^{\text{BCS}}_0$ as a function of the dimensionless parameter $T / T^{\text{BCS}}_c$, where $T^{\text{BCS}}_c$ is the critical temperature of the BCS theory and $\Delta^{\text{BCS}}_0 \equiv \Delta^{\text{BCS}} (T=0)$ is the zero-temperature gap parameter for the BCS theory \cite{bardeen1957theory}. We find that $|\Delta_1| = \sqrt{\overline{\Delta}_1 \Delta_1}$ is real, while $\sqrt{\overline{\Delta}_2 \Delta_2}$ is complex.
        
        The results presented in Fig. \ref{figure : gap solution} confirm that the gap parameters remain small in the vicinity of the critical temperature, thus justifying the expansion of the trace-log term in the effective action (\ref{effective action}).
        \subsection{Two-component complex Ginzburg-Landau model}
        We expand the effective action (\ref{effective action}) with respect to the combined fields $\Delta_\epsilon$ and $\Delta_\delta$. Note that the gap parameters shown in Fig. \ref{figure : gap solution} are obtained by truncating the gap equations at the order $\delta \epsilon$. However, truncating the effective action up to order $\delta \epsilon$ would not give the appropriate gap equations of order $\delta \epsilon$ because gap parameters $\Delta_\epsilon$ and $\Delta_\delta$ also scales as $\Delta_2 = \mathcal{O}(\delta) \Delta_1$ from Eq. (\ref{relation between delta 1 and delta 2}). Therefore, our procedure is to (i) expand the action with respect to $\Delta_\epsilon $ and $\Delta_\delta$ up to all orders, (ii) calculate the equations of motion with respect to $\overline{\Delta}_1$ and $\overline{\Delta}_2$, (iii) cut off the equations of motion at the first order $\delta\epsilon$, and (iv) find the corresponding effective action of the equations of motion. We also make the approximation that the dynamical terms are $\nabla_i \overline{\Delta}_\epsilon \nabla_i \Delta\epsilon \sim \nabla_i \overline{\Delta}_1 \nabla_i \Delta_1$ and $\nabla_i \overline{\Delta}_\delta \nabla_i \Delta_\delta \sim \nabla_i \overline{\Delta}_2 \nabla_i \Delta_2$.

        Step (i): We expand the effective action to all orders, which yields
        \begin{eqnarray}\label{final effective action}
            \!\!S_{\text{eff}} \!\!&=&\!\! \int_0^{1/T} d \tau\int d^3 r ~ \alpha_1 \nabla_i \overline{\Delta}_1\nabla_i \Delta_1 + r_1 \overline{\Delta}_\epsilon \Delta_\epsilon + u_1 (\overline{\Delta}_\epsilon \Delta_\epsilon)^2 +\alpha_2 \nabla_i \overline{\Delta}_2\nabla_i \Delta_2 + r_2 \overline{\Delta}_\delta\Delta_\delta \nonumber\\
            &&+ u_2 (\overline{\Delta}_\delta \Delta_\delta)^2-\frac{1}{g}\left[\overline{\Delta}_1\Delta_1 -i \frac{\epsilon }{\delta} \overline{\Delta}_2 \Delta_2 - i \epsilon \left(\overline{\Delta}_1\Delta_2+\overline{\Delta}_2\Delta_1\right)\right],
        \end{eqnarray}
        where each coefficients are given by $\alpha_i  (T) \propto \nu_i/T^2$, $u_i (T) \propto \nu_i/T^3$ and
        \begin{eqnarray}\label{Eq. Definition of r}
             r_i(T) &=&\frac{T}{V}\sum_{n,\vec{p}}\frac{1}{(\omega_n^2 + (\epsilon^{(i)}_{\vec{p}})^2 )}= \frac{\nu_i}{2} \int_0^{\frac{\omega_D}{2T}}dx \frac{\tanh(x)}{x}\equiv  \nu_i r(T) \sim \frac{\nu_i }{2} \log \left(\frac{2 e^{\gamma}\omega_D}{\pi T} \right),
        \end{eqnarray}
        as $T\rightarrow 0$. Here we have defined $r(T) :=  \frac{1}{2} \int_0^{\frac{\omega_D}{2T}}dx \frac{\tanh(x)}{x}$.

        Step (ii): Let us take a functional derivative of the potential of the complex action (\ref{final effective action}).  
        \begin{eqnarray}
            \frac{\delta S_{\text{eff}}}{\delta \overline{\Delta}_1} &=& r_1 \Delta_\epsilon + 2 u_1 \left(\overline{\Delta}_\epsilon \Delta_\epsilon\right)\Delta_\epsilon + \delta r_2  \Delta_\delta + 2 \delta u_2 \left(\overline{\Delta}_\delta \Delta_\delta\right)\Delta_\delta -\frac{1}{g} \Delta_1 + i \frac{1}{g} \epsilon \Delta_2 ,\\
            \frac{\delta S_{\text{eff}}}{\delta \overline{\Delta}_2} &=& r_2 \Delta_\delta + 2 u_2 \left(\overline{\Delta}_\delta \Delta_\delta\right)\Delta_\delta -i \epsilon r_1   \Delta_\epsilon -i \epsilon 2 u_1 \left(\overline{\Delta}_\epsilon \Delta_\epsilon\right)\Delta_\epsilon +i \frac{1}{g} \frac{\epsilon}{\delta}\Delta_2 + i \frac{1}{g} \epsilon \Delta_1,
        \end{eqnarray}
        where $\Delta_\epsilon = \Delta_1 - i \epsilon \Delta_2$ and $\Delta_\delta = \Delta_2 + \delta \Delta_1$. 
        
        Step (iii): Using the scaling relation $\Delta_2 = \mathcal{O}(\delta) \Delta_1$ from Eq. (\ref{relation between delta 1 and delta 2}), we can truncate the equations of motion:
        \begin{eqnarray}
            0&=&(r_1 - \frac{1}{g}+ \delta^2 r_2)\Delta_1 + 2 u_1 (\overline{\Delta}_1\Delta_1)\Delta_1+ r_2 \delta \Delta_2,\quad\quad\label{equations of motion}\\
            0&=&(-i \epsilon r_1 + \delta \{r_2 - \frac{1}{g}\frac{\epsilon}{i \delta}\})\Delta_1+ (r_2 - \frac{1}{g}\frac{\epsilon}{i \delta})\Delta_2.\quad\quad\label{equations of motion 2}
        \end{eqnarray}
        To be consistent, the above equations must match with the gap equations (\ref{Eq. Gap equation}). The consistency check is given in the \ref{Appendix: Consistancy check}, and indeed, we find two equations to match.
        
        Step (iv): The corresponding complex action of the equations of motion (\ref{equations of motion}) and (\ref{equations of motion 2}) is
        \begin{eqnarray}\label{Complex Ginzburg Landau}
            S_{\text{eff}} &=& \int_0^{1/T} d \tau\int d^3 r ~ \alpha_1 \nabla_i \overline{\Delta}_1 \nabla_i \Delta_1 + \left(r_1 - \frac{1}{g}+r_2 \delta^2\right)\overline{\Delta}_1 \Delta_1+ u_1 \left(\overline{\Delta}_1 \Delta_1\right)^2\\
            &&+ \alpha_2 \nabla_i \overline{\Delta}_2 \nabla_i \Delta_2 + \left(r_2 - \frac{1}{g}\frac{\epsilon}{i \delta}\right)\overline{\Delta}_2 \Delta_2+\left( -i \epsilon r_1 + \delta \left\{r_2 - \frac{1}{g}\frac{\epsilon}{i \delta}\right\}\right)\left(\overline{\Delta}_1 \Delta_2+\overline{\Delta}_2 \Delta_1\right)\nonumber
        \end{eqnarray}
        Notice that the above action decouples to the standard Ginzburg-Landau model and the free field theory by taking $\mu =0$ where $\epsilon = \mu / \gamma $ and $\delta = \mu / g$. Therefore, we will refer to the non-Hermitian scalar field theory above as the \textit{two-component complex Ginzburg-Landau model}. 
\section{The breakdown of the Meissner effect at the zero exceptional point}\label{Section: Main result}
    In the previous section, we obtained a two-component complex Ginzburg-Landau model by perturbing with respect to $\epsilon$ and $\delta$. This effective action, given by Eq.~(\ref{Complex Ginzburg Landau}), has also been considered by us and other authors in the context of particle physics \cite{fring2022massive,mannheim2019goldstone}, where a breakdown of the Higgs mechanism occurs at a parameter limit known as the 0EP. This section presents our main result: the Meissner effect breaks down at the 0EP, while the gap parameters remain finite.

   To analyze the Meissner effect, we introduce a classical magnetic field through the minimal coupling $\nabla_{\vec{r}} \rightarrow  \nabla_{\vec{r}} + e \vec{A}$, where $\vec{A}$ is the gauge field, and add the kinetic term of the Magnetic field $S_{\text{Mag}} =\int~ \kappa (\nabla\times \vec{A})^2 $ to the effective action (\ref{effective action}), where the magnetic field is defined by $\vec{B} \equiv \nabla \times \vec{A}$. Taking a functional derivative of the action with respect to the gauge field $\vec{A}$ gives the equations of motion $\delta S_{\text{eff}} / \delta A_i  =0$. Inserting the constant solutions $\{\Delta_i,\overline{\Delta}_i\}$, $i\in \{1,2\}$ and acting with $\nabla \times $, we arrive at the extended London equation for the non-Hermitian two-band BCS model: $\nabla^2 \vec{B} = 2e^2 \alpha_1/\kappa \left(\overline{\Delta}_1 \Delta_1 +\frac{\alpha_2}{\alpha_1} \overline{\Delta}_2\Delta_2\right)\vec{B}$. We say that the Meissner effect is broken when the right-hand side of this equation is zero. 
    
    Inserting the solutions of the equations of motion of the effective action (\ref{Complex Ginzburg Landau}), which are given explicitly in the next section, we have 
    \begin{eqnarray}\label{London equation step 2}
        \nabla^2 \vec{B} &=& \frac{2e^2 \alpha_2}{\kappa} \overline{\Delta}_1\Delta_1\vec{B}  \left[1 +\frac{\nu_2}{\nu_1}\left(-\delta + \frac{\epsilon}{\gamma} \frac{r_1}{r_2^2 + \frac{1}{\gamma^2}}+ i \epsilon \frac{r_1 r_2}{r_2^2 + \frac{1}{\gamma^2}}\right)^2 \right].
    \end{eqnarray} 
    The right-hand side of the London equation is often called the London penetration depth. Notice that our London equation differs from the conventional London equation, where the penetration depth can be complex. We leave the investigation of the consequence of the complex penetration depth for future study. Still, we note that the London equation with complex penetration depth has recently been observed in the context of the parity-breaking superconductor \cite{staalhammar2023emergent}. Although our motivation is the superconductor in the open quantum system, which differs from their consideration, the resulting London equation bears similarity with our London equation.
    
    The quantity in the square bracket of Eq.(\ref{London equation step 2}) vanishes when the following two equations are simultaneously satisfied:
    \begin{eqnarray}
        -\delta + \frac{\epsilon}{\gamma} \frac{r_1}{r_2^2 + \frac{1}{\gamma^2}}= 0 ,~~
        1 - \frac{\nu_2}{\nu_1}\epsilon^2 \left(\frac{r_1 r_2}{r_2^2 + \frac{1}{\gamma^2}}\right)^2 =0.
    \end{eqnarray}
    Two equations can be combined into one, giving $1 = \nu_2  \delta^2 \gamma^2 r_2^2 / \nu_1 $. Writing the explicit form of $r_2$ given in Eq. (\ref{Eq. Definition of r}), we find $\frac{1}{\nu_1}\frac{T}{V}\sum_{n,\vec{p}}1/(\omega_n^2 + (\epsilon^{(1)}_{\vec{p}})^2 )= (\epsilon/g \delta^2)( \sqrt{\nu_1}/  \nu_2 \sqrt{\nu_2})$.
    We can solve this equation in the standard way and find the temperature $T_{\text{EP}}$ at which the above equations are satisfied:
    \begin{eqnarray}\label{exceptional temperature}
        \frac{T_{\text{EP}}}{T_{c}^{\text{BCS}}} = e^{ \frac{1}{g\nu_1}-\frac{1}{g\nu_1} \left(\frac{\epsilon}{\delta^2}\frac{\nu_1 \sqrt{\nu_1}}{\nu_2 \sqrt{\nu_2}}\right)},
    \end{eqnarray}
    where $T_{c}^{\text{BCS}}$ is the BCS critical temperature. We refer to the above temperature as the \textit{exceptional temperature}.

    The exceptional temperatures $T_{\text{EP}}$ are indicated as vertical lines in Fig. \ref{figure : gap solution} for $1/g \nu_1 = 1/0.39$, $\nu_1 = \sqrt{11.7}$, $\nu_2 = \sqrt{9.47}$, and $\delta = 1/10$, with three values of the parameters $\epsilon=1/30$, $\epsilon = 1/110$, and $\epsilon=1/120$. The gap parameters ${\Delta_1 , \Delta_2}$ are only plotted for $\epsilon = 1/110$, as the plots for $\epsilon = 1/30$ and $\epsilon = 1/120$ are indistinguishable from that of $\epsilon = 1/110$. Importantly, the gap parameters $\Delta_1$ and $\Delta_2$ remain finite at $T_{\text{EP}}$, indicating that the Meissner effect breaks down at this exceptional temperature while the gap parameters remain finite.

    In the next section, we will show that the exceptional temperature $T_{\text{EP}}$ is equivalent to the 0EP of the two-component complex Ginzburg-Landau model (\ref{Complex Ginzburg Landau}). However, before proceeding, we would like to note a few additional features of the exceptional temperature $T_{\text{EP}}$.

    The exceptional temperature $T_\text{EP}$ starts near zero when $\epsilon = 1/30$, approaches the BCS critical temperature $T^{\text{BCS}}_{c}$ when $\epsilon = 1/110$, and surpasses $T^{\text{BCS}}_{c}$ when $\epsilon = 1/120$, where the breakdown of the Meissner effect is trivial due to the zero gap parameters (i.e. zero penetration depth) for $T>T^{\text{BCS}}_{c}$. This behavior indicates a restricted range of $\epsilon$ where the exceptional temperature $T_{\text{EP}}$ is detectable before it is obscured by $T^{\text{BCS}}_{c}$.

    The reason for this mechanism can be seen from the explicit form of the exceptional temperature (\ref{exceptional temperature}), which is equal to the BCS critical temperature when the exponent of the equation (\ref{exceptional temperature}) is zero. Therefore, we identify the detectable region of the exceptional temperature to be $\left(\mu/g\right)^2 \left(\nu_1/\nu_2\right)^{3/2} < \mu/ \gamma$, and the undetectable regions to be $\left(\mu/g\right)^2 \left(\nu_1/\nu_2\right)^{3/2} > \mu /\gamma$. Hence, we conclude that a careful balance of the three parameters $\mu$, $g$, and $\gamma$ is necessary to realize the exceptional temperature within the detectable range $T_{\text{EP}}<T^{\text{BCS}}_{c}$.

\subsection {Connection to the zero exceptional point}
    Let us analyze the spontaneous symmetry breaking of the effective action (\ref{Complex Ginzburg Landau}). A similar form of this action was analyzed in the particle-physics context by several authors \cite{fring2022massive,mannheim2019goldstone,alexandre2017symmetries,alexandre2018spontaneous,begun2021phase}, where it was discovered that at the 0EP \cite{fring2020goldstone}, the action becomes non-diagonalizable, the gauge masses vanish \cite{fring2022massive,mannheim2019goldstone} and the t'Hooft Polyakov monopole masses vanish \cite{fring2020t}. 

    To simplify the effective action given by Eq. (\ref{Complex Ginzburg Landau}), we define  $a_1 \equiv r_1 - \frac{1}{g}+r_2 \delta^2$, $a_2 \equiv r_2 -  \frac{1}{i \gamma}$, $b \equiv -i \epsilon r_1 + \delta r_2 + i \epsilon / g $. We further simplify the effective action by decomposing the first gap parameter: $\Delta_1 = \phi_1 + i \chi_1$, $\overline{\Delta}_1 = \phi_1- i \chi_1$, where $\phi_1 , \chi_1 \in \mathbb{R}$. This decomposition is justified by our approximation $\delta \epsilon \ll 1$ where the solution $\overline{\Delta}_1 \Delta_1$ is real (see Fig. \ref{figure : gap solution}). The resulting effective action is
        \begin{eqnarray}\label{action with zero EP}
            S_{\text{eff}} &=& \int_0^{1/T} d \tau\int d^3 r ~ \alpha_1 \partial_\mu \phi_1 \partial_\mu \phi_1 +\alpha_1 \partial_\mu \chi_1 \partial_\mu \chi_1 + \alpha_2 \partial_\mu \overline{\Delta}_2 \partial_\mu \Delta_2\\
            &&+  a_1 \left(\phi_1^2 + \chi_1^2\right)+u_1 \left(\phi_1^2 + \chi^2_1\right)^2+  a_2 \overline{\Delta}_2 \Delta_2\nonumber\\
            && + b \phi_1 \left( \Delta_2 +\overline{\Delta}_2\right)+i b \chi_1 \left( \Delta_2-\overline{\Delta}_2 \right). \nonumber
        \end{eqnarray}
        Recall the coefficients of the kinetic term and the fourth-order terms are defined as $\alpha_i  (T) \propto \nu_i/T^2$, $u_i (T) \propto \nu_i/T^3$.
        
        The solutions of equations of motion of this action are 
        \begin{eqnarray}\label{simple vacuume solutions}
            &({\phi_1^{}}^2 + {\chi_1^{}}^2) = - \frac{1}{2u_1} \left( a_1 - \frac{b^2}{a_2}\right),\\
            &\Delta_2^{} = - \frac{b}{a_2} ({\phi_1^{}} + i {\chi_1^{}}),~\overline{\Delta}_2^{} = - \frac{b}{a_2} ({\phi_1^{}} - i {\chi_1^{}}). 
        \end{eqnarray}
        Two equations coincide with the equations (\ref{equations of motion}) and (\ref{equations of motion 2}) up to order $\mathcal{O}(\delta^5)$.
    
    The above action is $U(1) $ symmetric with respect to the transformation $\Delta_1 \rightarrow e^{i\theta}\Delta_1$ and $\Delta_1 \rightarrow e^{-i\theta}\Delta_1$ with real parameter $\theta\in\mathbb{R}$. This symmetry is spontaneously broken when Taylor expands the above action around the solutions (\ref{simple vacuume solutions}), leading to the massless fields due to the Goldstone theorem. The solution (\ref{simple vacuume solutions}) is also $U(1) $ symmetric, which allows us to choose a specific solution $\phi_1^{}=0$ and $(\chi_1^{})^2 =- \left( a_1 - b^2/a_2\right)/2u_1 $. Expanding the effective action around this solution, one finds 
        \begin{eqnarray}
            S_{\text{eff}} &=& \int_0^{1/T} d \tau\int d^3 r ~\alpha_1 \partial_\mu \phi_1 \partial_\mu \phi_1 +\alpha_1 \partial_\mu \chi_1 \partial_\mu \chi_1 + \alpha_2 \partial_\mu \overline{\Delta}_2 \partial_\mu \Delta_2\\
            && +\left(\begin{array}{cc}
                \phi_1 &\overline{\Delta}_2  
            \end{array}\right) \left(\begin{array}{cc}
                 \frac{b^2}{a_2} & b  \\
                  b&  a_2  
            \end{array}\right)\left(\begin{array}{c}
                 \phi_1  \\
                 \Delta_2 
            \end{array}\right)+\dots , \nonumber
        \end{eqnarray}
        where ellipsis contains terms with $\chi_1$ and $\Delta_2$. The $2\times 2$ matrix is not yet a mass matrix of the field theory because of the unbalanced scaling difference in the kinetic term. More explicitly, let us denote $\Box=\partial_t^2 + \nabla_{\vec{r}}^2$ and rewrite the quadratic order of $\phi_1 $ and $\Delta_2$:
        \begin{eqnarray}
            S_{\text{eff}} &=& \int_0^{1/T} d \tau\int d^3 r ~ \left(\begin{array}{cc}
                \phi_1 &\overline{\Delta}_2  
            \end{array}\right)\left[\left(\begin{array}{cc}
                -\alpha_1 \Box &0  \\
                 0& -\alpha_2 \Box
            \end{array}\right)+ \left(\begin{array}{cc}
                 \frac{b^2}{a_2} & b  \\
                  b&  a_2  
            \end{array}\right)\right]\left(\begin{array}{c}
                 \phi_1  \\
                 \Delta_2 
            \end{array}\right)+\dots \\
            &=&  \int_0^{1/T} d \tau\int d^3 r ~ \frac{1}{T^2}\left(\begin{array}{cc}
                \phi_1 &\overline{\Delta}_2  
            \end{array}\right)\left(\begin{array}{cc}
                \nu_1 &0  \\
                0 &\nu_2 
            \end{array}\right)\left[-\Box \mathbb{I}+ \left(\begin{array}{cc}
                 \frac{1}{\nu_1}\frac{b^2}{a_2} & \frac{1}{\nu_1}b  \\
                  \frac{1}{\nu_2}b& \frac{1}{\nu_2} a_2  
            \end{array}\right)\right]\left(\begin{array}{c}
                 \phi_1  \\
                 \Delta_2 
            \end{array}\right)+\dots\nonumber
        \end{eqnarray}
        From the above action, the non-Hermitian mass matrix can be identified as
        \begin{eqnarray}
            M := \left(\begin{array}{cc}
                 \frac{1}{\nu_1}\frac{b^2}{a_2} & \frac{1}{\nu_1}b  \\
                  \frac{1}{\nu_2}b& \frac{1}{\nu_2} a_2  
            \end{array}\right)= S \left(\begin{array}{cc}
                0 & 0 \\
                0 & \frac{a_2}{\nu_2 }  +
                \frac{b^2}{\nu_1 a_2} 
            \end{array}\right)S^{-1}
        \end{eqnarray}
        where $S$ is the similarity matrix that diagonalizes the non-Hermitian mass matrix. The detail of the similarity matrix $S$ is omitted as it is not important for our current discussion. The Goldstone theorem ensures that one of the eigenvalues of the non-Hermitian mass matrix is zero due to the spontaneous symmetry breaking of the $U(1)$ symmetry. The non-Hermitian mass matrix becomes non-diagonalizable, and the similarity matrix $S$ becomes ill-defined when each parameter satisfies the equation $\nu_2 b^2 /a_2^2 \nu_1 +1=0 $. Recalling that $a_2 \equiv r_2 -  \frac{1}{i \gamma}$, $b \equiv -i \epsilon r_1 + \delta r_2 + i \epsilon / g $, we obtain the condition for non-diagonalizability of the mass matrix $M$ as a function of the parameters $\{\delta , \epsilon ,r_1 , r_2\}$:
        \begin{eqnarray}
            1+ \frac{\nu_1}{\nu_2}  \left(\frac{b}{a_2}\right)^2 =0 \iff 1 +\frac{\nu_1}{\nu_2}\left(-\delta + \frac{\epsilon}{\gamma} \frac{r_1}{r_2^2 + \frac{1}{\gamma^2}}+ i \epsilon \frac{r_1 r_2}{r_2^2 + \frac{1}{\gamma^2}}\right)^2   =0      .   
        \end{eqnarray}
        The above quantity is precisely the one that appears on the right-hand side of the extended London equation (\ref{London equation step 2}). Therefore, the mass matrix $M$ is non-diagonalizable at the exceptional temperature given by Eq. (\ref{exceptional temperature}). We can infer that the Meissner effect breaks down at the 0EP, while the gap parameters remain finite.

        Finally, we comment on the connection to the $\mathcal{PT}$ symmetry. The mass matrix considered previously in the particle physics context is equivalent to our mass matrix if the parameter $a_2$ is real and $b$ is imaginary \cite{fring2022massive}. In this case, the model has a simple anti-linear symmetry on the fields:
        \begin{eqnarray}
            \mathcal{PT} : \phi_1 \rightarrow\phi_1 ,\quad  \mathcal{PT} :   \Delta_2\rightarrow - \Delta_2 , \quad \mathcal{PT} :   \overline{\Delta}_2\rightarrow - \overline{\Delta}_2
        \end{eqnarray}
        However, in our case, the parameters $a_2$ and $b$ are both non-trivial complex functions of temperature. Therefore, it is non-trivial to identify whether a $\mathcal{PT}$ symmetry of the fields associated with the 0EP exists. We leave this as a future project. 
\section{Conclusion}
    We began by considering a two-band model weakly coupled to an external bath (see Fig. \ref{figure : schematic}), which can be reduced to a non-Hermitian two-band BCS model through the GKSL master equation. Employing the non-Hermitian mean field theory introduced in Ref. \cite{yamamoto2019theory} and path integral formalism, we derive a corresponding non-Hermitian scalar field theory that leads to a two-component complex Ginzburg-Landau model given in Eq. (\ref{Complex Ginzburg Landau}). By analyzing the Meissner effect of this model, we identified a temperature different from the critical temperature at which the Meissner effect breaks down while maintaining a finite gap parameter. We term this temperature the exceptional temperature and find it to be formally equivalent to the 0EP of the two-component complex Ginzburg-Landau model.

    In this work, we demonstrate the breakdown of the Meissner effect for a specific example, leaving the question of whether this is a general feature of a non-Hermitian many-body model with corresponding complex scalar field theory with continuous symmetry. The extended London equation Eq. (\ref{London equation step 2}) has acquired a complex part, which suggests a possible oscillatory magnetic field solution. In fact, such a solution was recently found in the context of parity-breaking superconductor \cite{staalhammar2023emergent}. We leave the search for such an oscillatory solution and the implication and possible connection to the 0EP for future projects. Furthermore, we observed a high sensitivity of the exceptional temperature to small variations in the coupling parameter $\epsilon$ compared to the gap parameter, motivating further research. Finally, while our current setup is a toy model, it provides insights into the role of EPs in open systems. However, realizing this setup experimentally is challenging, and more realistic systems remain to be explored in future work.

\section*{Acknowledgment}

We are grateful to Naomichi Hatano for the discussion on building the model, Tsuneya Yoshida for suggesting using the GKSL master equation, and Joshua Feinberg for fruitful discussions. TT is supported
by JSPS KAKENHI Grant Number 22KJ0752

\bibliographystyle{unsrt}
\bibliography{apssamp}

\appendix
\section{Gap equation}\label{Appendix: HS transformation}
    In this section, we derive the gap parameters plotted in Fig. \ref{figure : gap solution}, which are solutions to the gap equations $\delta S_{\text{eff}} / \delta \Delta_1 =0 $, $\delta S_{\text{eff}} / \delta \overline{\Delta}_1 =0 $, $\delta S_{\text{eff}} / \delta \Delta_2 =0 $ and $\delta S_{\text{eff}} / \delta \overline{\Delta}_2 =0 $.
    
    Taking the functional derivative of the effective action (\ref{effective action}), we find the following coupled gap equations
        \begin{eqnarray}
            \frac{\delta S_{\text{eff}}}{\delta \overline{\Delta}_1}&=& \frac{1}{g}  - \frac{T}{V}\sum_{n,\vec{p}}\frac{1}{\omega_n^2 + (\epsilon^{(1)}_{\vec{p}})^2 + \overline{\Delta}_\epsilon\Delta_\epsilon}- \frac{\mu}{g}\frac{\Delta_\delta}{\Delta_\epsilon}\frac{T}{V}\sum_{n,\vec{p}}\frac{1}{\omega_n^2 + (\epsilon^{(2)}_{\vec{p}})^2 + \overline{\Delta}_\delta\Delta_\delta}=0 , \label{gap equation 1 step 1}\\
            \frac{\delta S_{\text{eff}}}{\delta \overline{\Delta}_2} &=&\frac{1}{i\gamma}  - \frac{T}{V}\sum_{n,\vec{p}}\frac{1}{\omega_n^2 + (\epsilon^{(2)}_{\vec{p}})^2 + \overline{\Delta}_\delta \Delta_\delta}- \frac{\mu}{i\gamma}\frac{\Delta_\epsilon}{\Delta_\delta}\frac{T}{V}\sum_{n,\vec{p}}\frac{1}{\omega_n^2 + (\epsilon^{(1)}_{\vec{p}})^2+ \overline{\Delta}_\epsilon\Delta_\epsilon} =0 ,\quad\label{gap equation 2 step 1}
        \end{eqnarray}
        where $T$ and $V$ are the system's temperature and volume. Note that combined gap parameters $\{\Delta_\epsilon , \overline{\Delta}_\epsilon\}$ and $\{\Delta_\delta , \overline{\Delta}_\delta\}$ in the above equations are constant solutions (no dependence of space and time) with temperature dependence. Furthermore, solving the rest of the gap equations $\delta S_{\text{eff}}/ \delta \Delta_1 =0$ and $\delta S_{\text{eff}}/ \delta \Delta_2 =0$ result in the same equations as above but with fields $\Delta_1$ and $\Delta_2$ replaces with $\overline{\Delta}_1$ and $\overline{\Delta}_2$, respectively. 

        Next, notice that two equtions (\ref{gap equation 1 step 1}) and (\ref{gap equation 2 step 1}) can be combined and rewritten in a following form 
        \begin{eqnarray}
            0&=&\frac{1}{g}- \frac{T}{V}\sum_{n,\vec{p}}\frac{1}{\omega_n^2 + (\epsilon^{(1)}_{\vec{p}})^2 + \overline{\Delta}_\epsilon \Delta_\epsilon}- i \delta \epsilon \frac{T}{V}\sum_{n,\vec{p}}\frac{1}{\omega_n^2 + (\epsilon^{(1)}_{\vec{p}})^2 + \overline{\Delta}_\epsilon \Delta_\epsilon} + i \frac{\delta}{\gamma} \frac{\Delta_\delta}{\Delta_\epsilon}  ,\label{full gap equation 1}\\
            0&=&\frac{1}{i\gamma}- \frac{T}{V}\sum_{n,\vec{p}}\frac{1}{\omega_n^2 + (\epsilon^{(2)}_{\vec{p}})^2 + \overline{\Delta}_\delta\Delta_\delta} - i \delta \epsilon \frac{T}{V}\sum_{n,\vec{p}}\frac{1}{\omega_n^2 + (\epsilon^{(2)}_{\vec{p}})^2 + \overline{\Delta}_\delta\Delta_\delta} + i \frac{\epsilon}{g} \frac{\Delta_\epsilon}{\Delta_\delta}  .\label{full gap equation 2}
        \end{eqnarray}
        We define the critical temperature $T_c$ of our model to be such that the gap parameters $\Delta_\epsilon$ and $\Delta_\delta$ vanish. However, it is ambiguous whether the last terms in the equations (\ref{full gap equation 1}) and (\ref{full gap equation 2}) vanish or diverge. Therefore we make an Ansatz for $\Delta_\epsilon$ and $\Delta_\delta$ such that
        \begin{eqnarray}\label{relation between combined fields}
            \frac{\Delta_\delta}{ \Delta_\epsilon}(T=T_c) = A+ i B ~,~~~\frac{\overline{\Delta}_\delta }{ \overline{\Delta}_\epsilon}(T=T_c) = A+ i B,\nonumber\\
        \end{eqnarray}
        or equivalently 
        \begin{eqnarray}\label{relation between gap parameters}
            \Delta_2 (T_c) = \frac{A+ i B -\delta}{1+i \epsilon A - \epsilon B} \Delta_1 ( T_c),\quad
            \overline{\Delta}_2 (T_c) = \frac{A+ i B -\delta}{1+i \epsilon A - \epsilon B} \overline{\Delta}_1 (T_c),
        \end{eqnarray}
        for some constants $A$ and $B$, which will be determined by solving the above two equations (\ref{full gap equation 1}) and (\ref{full gap equation 2}). 
        
        Note that it seems natural to choose an Ansatz such as $\Delta_\delta / \Delta_\epsilon (T=T_c) = A+ i B$ and $\overline{\Delta}_\delta / \overline{\Delta}_\epsilon(T=T_c) = A- i B$. However, this Ansatz leads to the incompatible set of equations between $\{\delta S_{\text{eff}} / \delta \Delta_1 =0,\delta S_{\text{eff}} / \delta \Delta_2 =0\}$ and $\{\delta S_{\text{eff}} / \delta \overline{\Delta}_1 =0,\delta S_{\text{eff}} / \delta \overline{\Delta}_2 =0\}$. This is because the solution to the first set of equations is not the solution to the second set of equations.
        
        Next, let us begin by taking $T=T_c$ and using the relation $\frac{T}{V}\sum_{n,\vec{p}}1/(\omega_n^2 + (\epsilon^{(2)}_{\vec{p}})^2 ) =\frac{\nu_2}{\nu_1}\frac{T}{V} \sum_{n,\vec{p}}1/(\omega_n^2 + (\epsilon^{(1)}_{\vec{p}})^2 )$, where $\{\nu_1 ,\nu_2\}$ are the densities of state of the BCS Hamiltonian $H_{\text{BCS}} [c_1 , c_1^\dagger] $ and the free Hamiltonian $H_{\text{Free}} [c_2 , c_2^\dagger] $ approximated near the Fermi surface.  The two equations (\ref{full gap equation 1}) and (\ref{full gap equation 2}) are rewritten as
        \begin{eqnarray}
            0&=&\frac{1}{g}+ i \frac{\delta}{\gamma} \left(A+i B\right)- \frac{T}{V}\sum_{n,\vec{p}}\frac{1}{\omega_n^2 + (\epsilon^{(1)}_{\vec{p}})^2}- i \delta \epsilon \frac{T}{V}\sum_{n,\vec{p}}\frac{1}{\omega_n^2 + (\epsilon^{(1)}_{\vec{p}})^2 }   ,\label{reduced gap equation 1}\\
            0&=&\frac{\nu_1}{\nu_2}\left[\frac{1}{i\gamma}+ i \frac{\epsilon}{g} \frac{1}{A+i B}\right]- \frac{T}{V}\sum_{n,\vec{p}}\frac{1}{\omega_n^2 + (\epsilon^{(1)}_{\vec{p}})^2}- i \delta \epsilon  \frac{T}{V}\sum_{n,\vec{p}}\frac{1}{\omega_n^2 + (\epsilon^{(1)}_{\vec{p}})^2 }   .\label{reduced gap equation 2}
        \end{eqnarray}
        The first two terms of equations (\ref{reduced gap equation 1}) and (\ref{reduced gap equation 2}) should be equal to each other to solve the two equations simultaneously. This constraint implies that we need to find the solutions $A$ and $B$ such that 
        \begin{eqnarray}
            -2 A B \epsilon +A-\frac{B \epsilon \nu_1}{\delta \nu_2  } &=& 0,\label{equation for A and B 1}\\
            A^2 \epsilon +\frac{\nu_1\epsilon  (A-\delta )}{\delta  \nu_2 }-B^2 \epsilon +B &=& 0,\label{equation for A and B 2}
        \end{eqnarray}
        which can be solved exactly. However, the explicit forms are quite cumbersome. Therefore, we will only give the approximate forms in the next section. The gap equations (\ref{reduced gap equation 1}) and (\ref{reduced gap equation 2}) are now reduced to a single gap equation with the real and the imaginary parts.
        \begin{eqnarray}
            0&=&\frac{1}{g}(1-\epsilon B) - \frac{T}{V}\sum_{n,\vec{p}}\frac{1}{\omega_n^2 + (\epsilon^{(1)}_{\vec{p}})^2} + i \delta \epsilon \left[\frac{1}{g}\frac{A}{\delta}-\frac{T}{V}\sum_{n,\vec{p}}\frac{1}{\omega_n^2 + (\epsilon^{(1)}_{\vec{p}})^2} \right] .\label{reduced reduced gap equation}
        \end{eqnarray}
        If the real part of the above equation is satisfied, the imaginary part is reduced to 
        \begin{eqnarray}\label{region where gap equation holds}
            i\frac{1}{g}\delta \epsilon \left(\frac{A}{\delta}-1+\epsilon B\right).
        \end{eqnarray}
        This quantity does not vanish for the exact values of $A$ and $B$. Therefore, we will consider an approximation with respect to the parameters of our theory.
    \subsection{Approximate solution to the gap equation}\label{Section: Approximation Scheme 1}
        Let us assume that the coupling between the BCS Hamiltonian and the free theory is weaker than the loss rate with the external bath, $\mu \ll \gamma$. This assumption means we take $\epsilon= \mu / \gamma$ as the perturbative parameter. For the expansion with respect to $\epsilon$ to be consistent, we need to assume $\epsilon \ll \delta$.
        
\begin{figure}
            \centering
            \begin{minipage}[b]{0.6\textwidth}           \includegraphics[width=\textwidth]{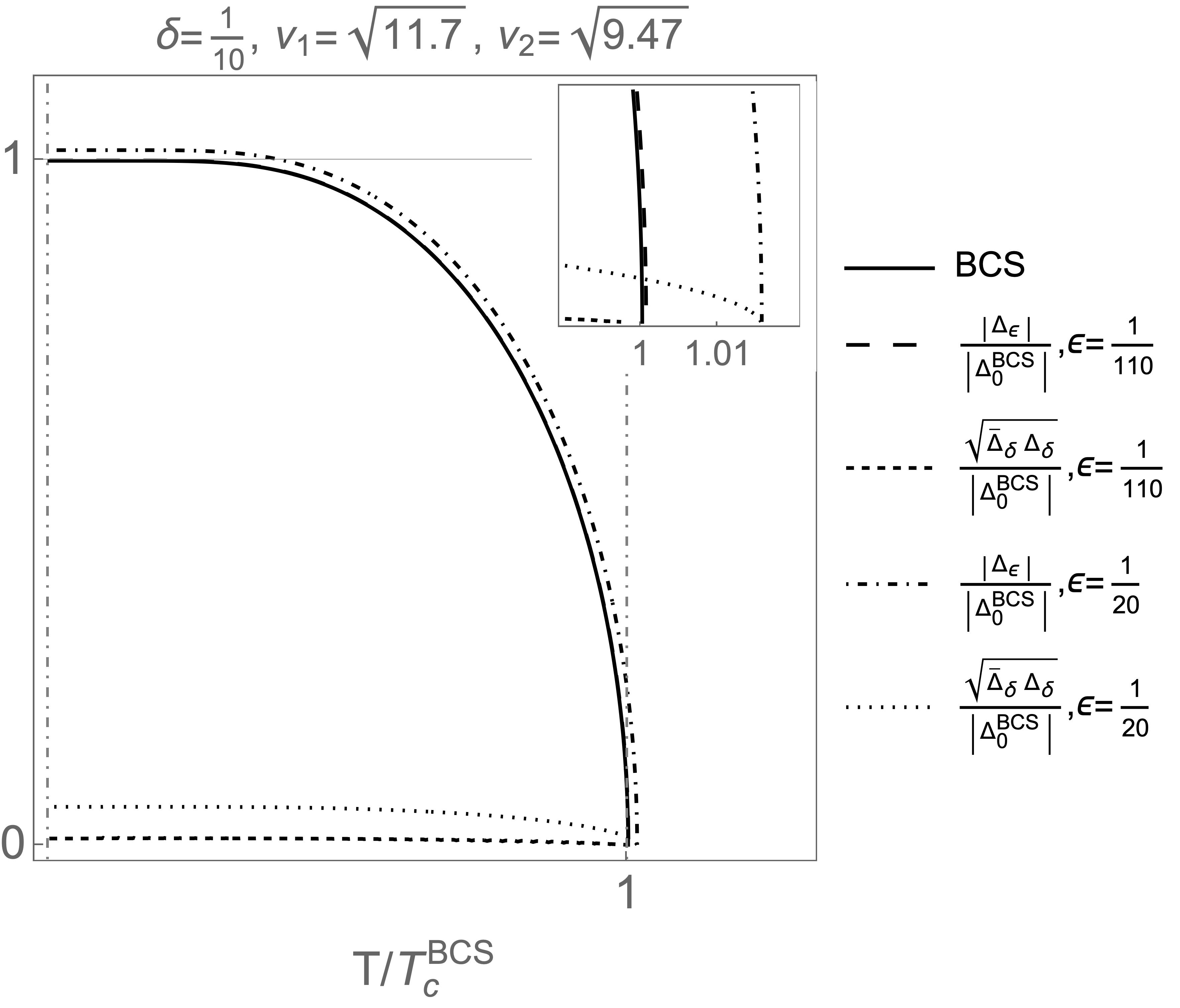}
            \end{minipage}
            \caption{A numerical plot of the gap parameters of the BCS model and our model as a function of temperature. They are obtained from the gap equations $\delta S_{\text{eff}}/ \delta \Delta_1 =0 $ and $\delta S_{\text{eff}}/ \delta \Delta_2 =0 $ with the approximation $\epsilon \ll \delta \ll 1$ up to, but not including, the order $\delta \epsilon$. Parameters are taken to be $1/g \nu_1 = 1/0.39$, $\nu_1  = \sqrt{11.7}$, $\nu_2 = \sqrt{9.47}$, and $\delta = 1/10$ with two values $\epsilon = 1/ 20$, $\epsilon = 1/ 110$. Solid and dotted lines are the real and imaginary parts of the solution.}
            \label{figure : gap solution 2}
        \end{figure}

        Expanding the explicit expressions of $A$ and $B$ with respect to $\epsilon$, we find 
        \begin{eqnarray}\label{approximation to A and B}
            A =\frac{\epsilon^2 }{\delta \nu^2} + \mathcal{O}(\epsilon^4) ~,~~~ B = \frac{\epsilon \nu_1}{\nu_2}+ \mathcal{O}(\epsilon^3),
        \end{eqnarray}
        where the expansion is valid with respect to the approximation $\epsilon \ll \delta \ll 1$. This is because each order of the series always appears in the form $\epsilon^n / \delta^m$ where $n>m$ are the positive integers. 
        The real part and the imaginary part of the gap equation (\ref{reduced reduced gap equation}) now take the following form 
        \small
        \begin{eqnarray}\label{reduced reduced reduced gap equation}
            0&=&\frac{1}{g}\left(1- \frac{\epsilon^2 \nu_1}{\nu_2}+ \mathcal{O}\left(\epsilon^{4}\right)\right) - \frac{T}{V}\sum_{n,\vec{p}}\frac{1}{\omega_n^2 + (\epsilon^{(1)}_{\vec{p}})^2}+ i \delta \epsilon\left[\frac{1}{g}\left(1+\mathcal{O}\left(\epsilon^{2}\right) \right) - \frac{T}{V}\sum_{n,\vec{p}}\frac{1}{\omega_n^2 + (\epsilon^{(1)}_{\vec{p}})^2}\right].\nonumber\\
        \end{eqnarray}
        \normalsize
        Notice that if the real part of the above equation vanishes, then the leading contribution of the complex part scales as $\mathcal{O}(\delta \epsilon^3)$. This observation implies that solving the real part of the above equation automatically ensures that the complex part can be ignored in our approximation. Solving the real part in a standard way, one finds the critical temperature of the system:
        \begin{eqnarray}\label{critical temperature}
            k_B T_c  \sim  2\hbar \omega_D \frac{e^{\gamma}}{\pi} e^{- \frac{1}{g \nu_1}\left(1 -\frac{\epsilon^2 \nu_1}{\nu_2}+\mathcal{O}(\delta^{4})\right)}  \text{ as }\frac{2 k_B T_c}{\hbar \omega_D} \rightarrow 0 .\nonumber\\
        \end{eqnarray}
        Assuming that the relation $\Delta_\delta / \Delta_\epsilon = \Delta_\epsilon (T=T_c) = A+ i B$ or equivalently the relation (\ref{relation between gap parameters}) holds away from the critical temperature, the gap equation (\ref{full gap equation 1}) now takes the following form 
        \begin{eqnarray}\label{reduced reduced gap eqation}
            \frac{1}{g}\left(1-\frac{\epsilon^2 \nu_1}{\nu_2}\right) -\frac{T}{V} \sum_{n,\vec{p}}\frac{1}{\omega_n^2 + (\epsilon^{(1)}_{\vec{p}})^2 + \overline{\Delta}_\epsilon \Delta_\epsilon }+ \mathcal{O}(\epsilon^{3}).\nonumber\\
        \end{eqnarray}
        This equation is solved numerically; the result is shown in Fig. \ref{figure : gap solution 2}. Since it is a real gap equation, the quantity $\overline{\Delta}_\epsilon \Delta_\epsilon$ is real. Therefore let us define $\sqrt{\overline{\Delta}_\epsilon \Delta_\epsilon} \equiv |\Delta_\epsilon| \in \mathbb{R}$. The second gap parameter is found by the relation (\ref{relation between combined fields}) and (\ref{approximation to A and B}), which gives $\overline{\Delta}_\delta \Delta_\delta = \left(\frac{\epsilon^2 \nu_1}{\delta \nu_2}+ i \frac{\epsilon \nu_1}{\nu_2}\right)^2 |\Delta_\epsilon|^2$. Keeping only the leading order contribution, we find $\sqrt{\overline{\Delta}_\epsilon \Delta_\epsilon}  = i \frac{\epsilon \nu_1}{\nu_2} |\Delta_\epsilon|$, where we find that the quantity $\sqrt{\overline{\Delta}_\epsilon \Delta_\epsilon}$ is pure imaginary. This result is plotted in Fig. \ref{figure : gap solution 2}. 
        Finally, let us find the gap equations for the parameters $\Delta_1$ and $\Delta_2$.
        \subsubsection{Solutions of gap parameters}\label{Section : relation between two gap parameters}
        Using the relation (\ref{relation between gap parameters}) and Eq. (\ref{approximation to A and B}) one finds 
        \begin{eqnarray}\label{relation between delta 1 and delta 2}
            \Delta_2 = \left[-\delta + i \frac{\epsilon \nu_1}{\nu_2}+ \mathcal{O}(\epsilon^2)\right] \Delta_1 \implies \Delta_\epsilon = \left[1+ i \delta \epsilon +\frac{\epsilon^2 \nu_1}{\nu_2}+\mathcal{O}(\epsilon^3)\right]\Delta_1 ,
        \end{eqnarray}
        near the critical temperature $T_c$. This allows us to expand the gap equation (\ref{reduced reduced gap eqation}) and find 
        \begin{eqnarray}\label{full gap equation for approximation 2}
            0&=&\frac{1}{g}\left(1-\frac{\epsilon^2 \nu_1}{\nu_2} + \mathcal{O}(\epsilon^4) \right)-\frac{T}{V} \sum_{n,\vec{p}}\frac{1}{\omega_n^2+ (\epsilon^{(1)}_{\vec{p}})^2 + |\Delta_1|^2}\nonumber\\
            &&-2 \left[i \delta \epsilon +\frac{\epsilon^2 \nu_1}{\nu_2}\right] |\Delta_1|^2   \frac{T}{V}\sum_{n,\vec{p}}\frac{1}{\left(\omega_n^2 + (\epsilon^{(1)}_{\vec{p}})^2 + |\Delta_1|^2\right)^{2}}+ \mathcal{O} (\delta
            ^4).
        \end{eqnarray}
        This equation has a complex part coming from the $\delta \epsilon $ term. Therefore, to find the complete picture, one needs to consider the real and the imaginary part of $\overline{\Delta}_1 \Delta_1$ separately. However, one more parameter $\delta$ is at our disposal to make a further approximation. Let us assume we cut off the above equation at order $\delta \epsilon$. Since we assume that $\epsilon \ll \delta$, the order $\epsilon^2$ must be smaller than $\delta \epsilon$. Therefore, the above equation is approximated to the standard BCS gap equation. 
        \begin{eqnarray}\label{gap equation for approximation 2}
            0=\frac{1}{g}\left(1+ \mathcal{O}(\epsilon^2) \right)- \frac{T}{V}\sum_{n,\vec{p}}\frac{1}{\omega_n^2 + (\epsilon^{(1)}_{\vec{p}})^2 + |\Delta_1|^2}.
        \end{eqnarray}
        We can conclude that the gap parameter $\overline{\Delta}_1 \Delta_1$ takes the real value when approximated to the order $\delta\epsilon$. This result is not surprising since weakening the approximation should eventually result in the standard BCS gap parameter. 
        The numerical solution to the above equation is plotted in Fig. \ref{figure : gap solution}. The second gap parameter is found by the relation (\ref{relation between delta 1 and delta 2}) 
        \begin{eqnarray}
            \sqrt{\overline{\Delta}_2 \Delta_2} = \left(-\delta + i \frac{\epsilon \nu_1}{\nu_2}\right) |\Delta_1|,
        \end{eqnarray}
        where we find that the quantity $\sqrt{\overline{\Delta}_2 \Delta_2}$ has a real and an imaginary part, which is also plotted in Fig. \ref{figure : gap solution}.
        
        Lastly, let us make a comment that a better approximation of the complex gap parameters $\{\Delta_\epsilon , \Delta_\delta\}$ or $\{\Delta_1, \Delta_2\}$ requires a higher-order correction to the gap equation. However, our main motivation is to expand the trace log term of the effective action (5) in the main text, which is only allowed when the gap parameters $\Delta_\epsilon$ and $\Delta_\delta$ are small near the critical temperature. Therefore, we do not need to consider the higher order correction to fulfill our main motivation.  
\section{Consistancy check}\label{Appendix: Consistancy check}
        From the second equation (\ref{equations of motion 2}), we find the relation between $\Delta_1$ and $\Delta_2$
        \begin{eqnarray}\label{vacuume solution 2 full}
            \frac{\Delta_2}{\Delta_1} (T) = \frac{i \epsilon r_1 - \delta \left\{r_2 - \frac{1}{g}\frac{\epsilon}{i \delta}\right\}}{r_2 -  \frac{1}{g}\frac{\epsilon}{i \delta}} .
        \end{eqnarray}
        This quantity can be evaluated at the critical temperature by using Eq. (\ref{reduced reduced gap eqation}), which gives $r_1 =  r_2 \nu_1/ \nu_2 = (1-\epsilon^2 \nu_1/ \nu_2 )/g$ and the above equation is reduced to $\Delta_2 (T_c) / \Delta_1 (T_c) = -\delta +i \epsilon \nu_1/ \nu_2 + \mathcal{O}(\epsilon^2)$. This approximation agrees with the relation obtained via the microscopic approach (\ref{relation between delta 1 and delta 2}). 
        
        The first equation can be solved by inserting the solution (\ref{vacuume solution 2 full}) to the Eq. (\ref{equations of motion}):
        \begin{eqnarray}\label{vacuume solution}
            \overline{\Delta}_1 \Delta_1 = - \frac{1}{2u_1}\left[r_1 - \frac{1}{g}+\delta^2 r_2 + \delta r_2  \frac{\Delta_2}{\Delta_1}\right]\rightarrow - \frac{1}{g2u_1}\left[i \delta \epsilon + \mathcal{O}(\epsilon^2)\right] \text{ as }T\rightarrow T_c.
        \end{eqnarray}
        In the previous section, we have cut off the gap equation at the order $\delta\epsilon$ to obtain the real gap parameter. This implies that the quantity $\overline{\Delta}_1 \Delta_1$ vanishes at the critical temperature. Therefore, we have shown the consistency between the microscopic and macroscopic theories up to the order $\delta \epsilon$.

\end{document}